\documentclass[conference,a4paper]{IEEEtran}
\IEEEoverridecommandlockouts

\usepackage{cite}
\usepackage{amsmath,amssymb,amsfonts}
\usepackage{algorithmic}
\usepackage{graphicx}
\usepackage{textcomp}
\usepackage{xcolor}
\usepackage{array}
\usepackage{booktabs} 
\usepackage{newtxtext} 
\usepackage{color}
\usepackage{hyperref}
\usepackage{url}
\usepackage{tabularx}
\usepackage[table]{xcolor} 
\usepackage{float}
\usepackage{booktabs}
\usepackage{multirow}
\usepackage{cleveref}
\usepackage{mathrsfs} 
\usepackage{graphicx} 
\usepackage{subcaption} 
\usepackage{makecell}

\usepackage{pifont}
\usepackage{booktabs} 
\renewcommand{\arraystretch}{1.2}
\def\BibTeX{{\rm B\kern-.05em{\sc i\kern-.025em b}\kern-.08em
    T\kern-.1667em\lower.7ex\hbox{E}\kern-.125emX}}
    
\begin{document}

\title{CoverAssert: Iterative LLM Assertion Generation Driven by Functional Coverage via Syntax-Semantic Representations
\thanks{\textdagger~ Corresponding Author
\\This paper has been published in the 29th Design, Automation and Test in Europe Conference (DATE 2026), April 20-22, 2026, Verona, Italy.
}
}

\author{
\IEEEauthorblockN{
Yonghao Wang\textsuperscript{1},
Yang Yin\textsuperscript{1},
Hongqin Lyu\textsuperscript{1,2}, 
Jiaxin Zhou\textsuperscript{3},
Zhiteng Chao\textsuperscript{1},
Mingyu
Shi\textsuperscript{4},
Wenchao
Ding\textsuperscript{2},
\\
Yunlin
Du\textsuperscript{5},
Jing
Ye\textsuperscript{1,2},
Tiancheng
Wang\textsuperscript{1\textdagger}
and
Huawei Li\textsuperscript{1,2\textdagger}}
\IEEEauthorblockA{\textsuperscript{1}State Key Lab of Processors, Institute of Computing Technology, CAS, Beijing, China}
\IEEEauthorblockA{\textsuperscript{2}University of Chinese Academy of Sciences, Beijing, China}
\IEEEauthorblockA{\textsuperscript{3}Beijing Normal University, China; \textsuperscript{4}Nanjing University, China; \textsuperscript{5}University of Newcastle, Australia}
\IEEEauthorblockA{\{wangyonghao22, yinyang22, lvhongqin24\}@mails.ucas.ac.cn, zhoujiaxin@mail.bnu.edu.cn,
\\
chaozhiteng@ict.ac.cn, mingyu.shi@smail.nju.edu.cn,
dingwenchaohd@gmail.com,
\\
Yunlin.Du@uon.edu.au,
\{yejing, wangtiancheng, lihuawei\}@ict.ac.cn}
}

\maketitle

\begin{abstract}
LLMs can generate SystemVerilog assertions (SVAs) from natural language specs, but single-pass outputs often lack functional coverage due to limited IC design understanding. We propose CoverAssert, an iterative framework that clusters semantic and AST-based structural features of assertions, maps them to specifications, and uses functional coverage feedback to guide LLMs in prioritizing uncovered points. Experiments on four open-source designs show that integrating CoverAssert with AssertLLM and Spec2Assertion improves average improvements of 9.57\% in branch coverage, 9.64\% in statement coverage, and 15.69\% in toggle coverage.
\end{abstract}


\section{Introduction}

Functional verification is essential in IC design, ensuring RTL code meets architectural specifications. Assertion-Based Verification (ABV) improves visibility and cuts debugging time by up to 50\% \cite{1,2}, but translating ambiguous specs into accurate SVAs is slow and difficult \cite{4,6}. LLM-based methods can generate assertions from specs or RTL \cite{10,11,12,14,20,6,22}, yet comprehensive functional coverage remains challenging.

A key challenge is accurately linking functional intent in specifications to assertions. High similarity among assertion codes and overlapping signal names across modules often lead LLMs to mistake lexical resemblance for true semantics, making it difficult to identify uncovered functional points and provide refinement feedback.

\begin{figure}[h]
  \centering
  \includegraphics[width=1\linewidth]{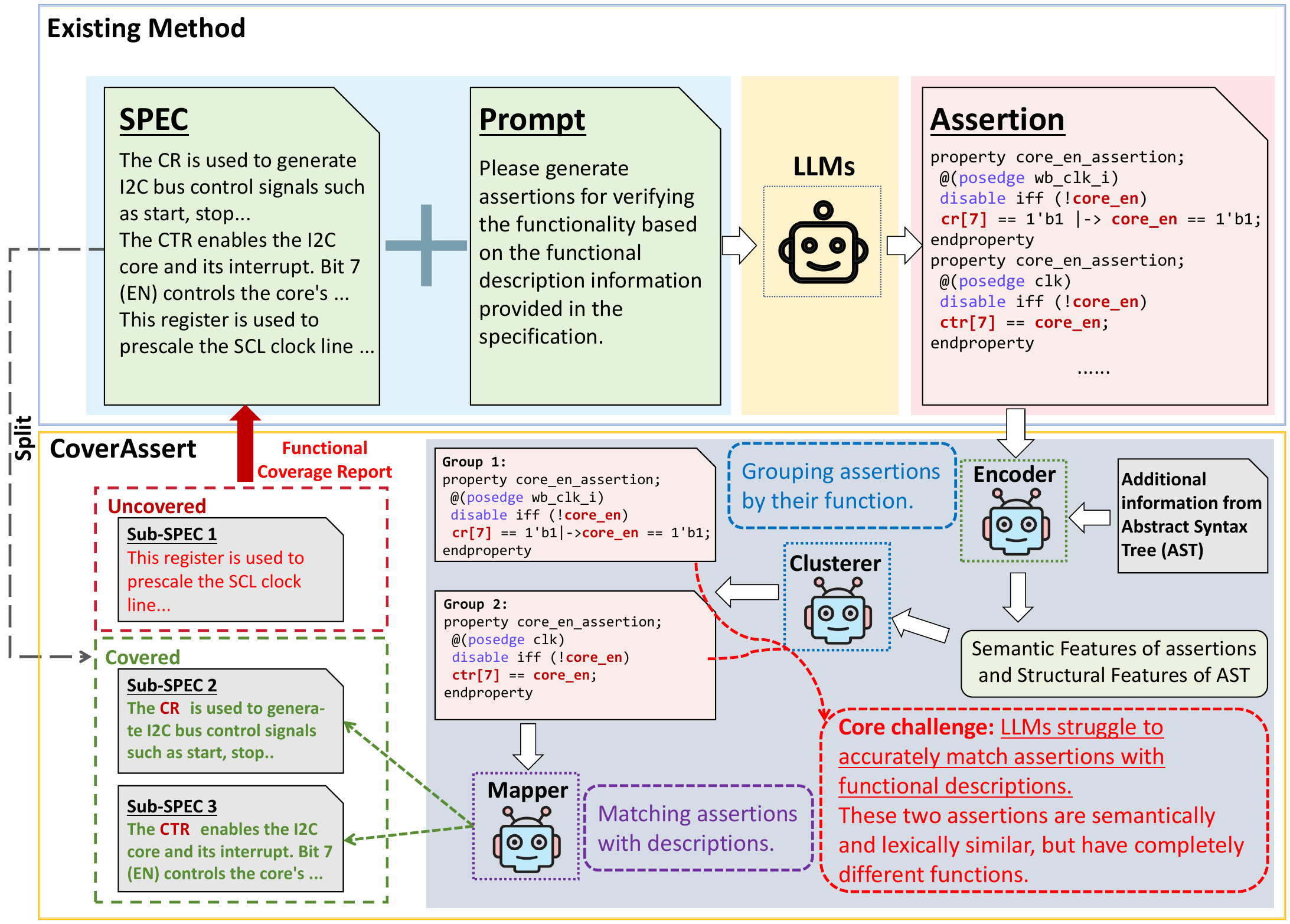}
  \caption{Our proposed framework, CoverAssert, directly enhances assertion generation by accurately identifying uncovered functional descriptions in the specification and providing feedback to prioritize the completion of uncovered functional points, thereby improving overall verification coverage.}
  \label{fig:intro}
\end{figure}

To address this, we propose CoverAssert, a lightweight feedback framework for LLM-based assertion generation. It iteratively identifies uncovered functions, enabling precise coverage analysis and guiding assertion refinement.

The contributions of this paper are summarized as follows:

\begin{enumerate}[]
\item We propose CoverAssert, a lightweight LLM-based iterative SVA framework that evaluates functional coverage via syntax–semantic assertion representations—the first to introduce coverage feedback into LLM assertion generation, compatible with existing methods.

\item CoverAssert fuses semantic embeddings with variable positions in syntax trees, enabling precise matching between assertions and functional descriptions.

\item Integrated with two SOTA methods, CoverAssert improved branch, statement, and toggle coverage by 9.5\%, 9.64\%, and 15.69\% on four open-source circuits.
\end{enumerate}

\section{Background}

The early work of automated hardware assertion generation proposed by Rahul Kande et al. \cite{12}, who pioneered the application of LLMs to generate assertions. AssertLLM by Fang et al. \cite{10} was a milestone, handling full specification files to produce detailed SVAs for all architectural signals. More recently, Wu et al. \cite{14} introduced Spec2Assertion, using progressive regularization and Chain-of-Thought prompting to generate high-quality assertions directly from specification.


\section{Framework of CoverAssert}

\begin{figure}[h]
  \centering
  \includegraphics[width=1\linewidth]{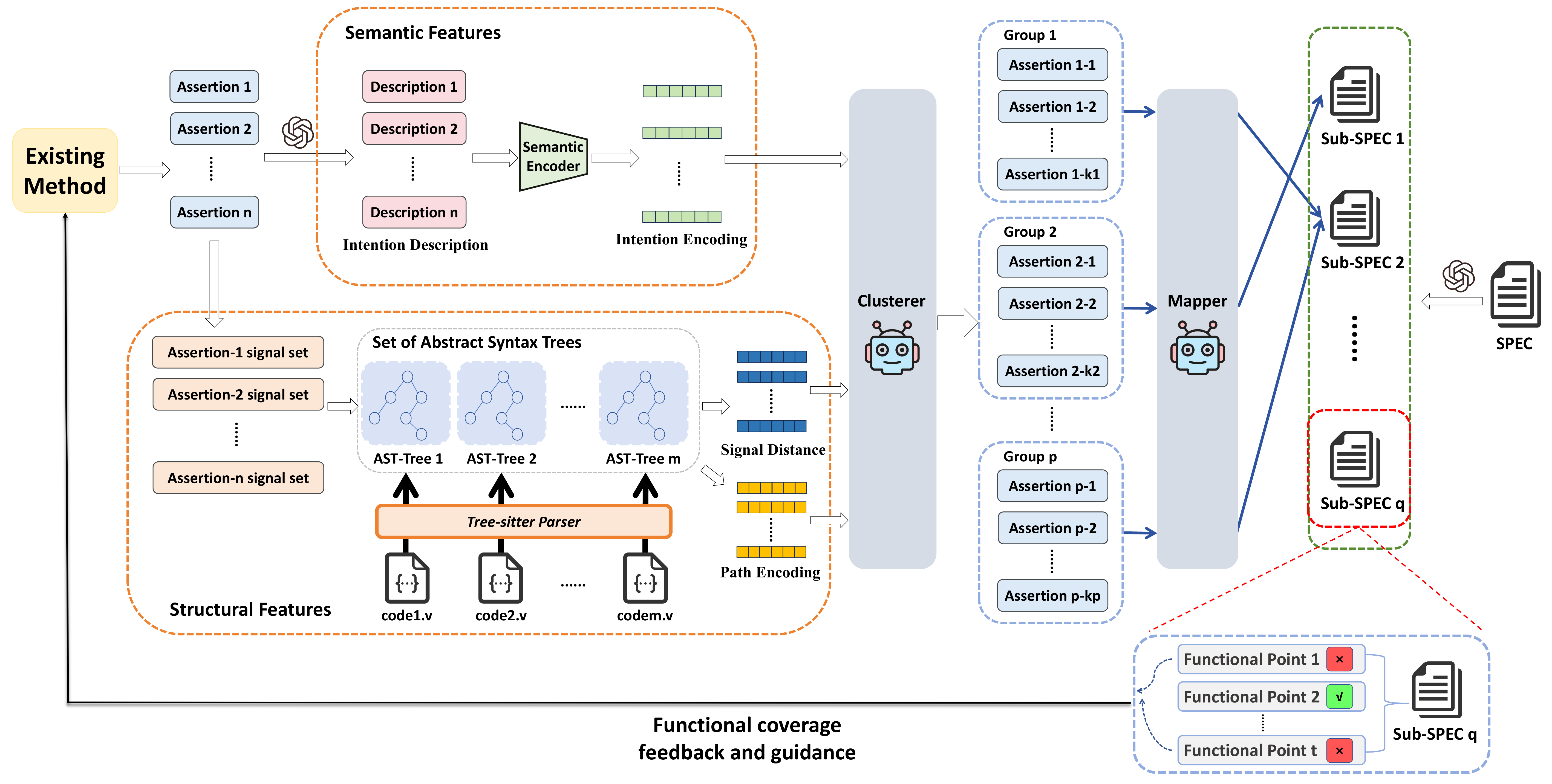}
  \caption{The CoverAssert framework integrates with existing assertion generation methods.}
  \label{fig:framework}
\end{figure}

We propose CoverAssert, a functional coverage–guided assertion generation framework. Its six-module workflow is shown in Fig. \ref{fig:framework}.

\subsubsection{\textbf{Semantic Feature Extraction}}

High syntactic similarity limits direct embeddings, so we first use an LLM (e.g., ChatGPT-4o) to extract intent, and then encode them using Qwen3-Embedding into vectors $T$ for semantic discrimination.

\subsubsection{\textbf{Signal Structural Features Extraction}}

For each assertion pair, signals are extracted and their pairwise longest common ancestor (LCA) distances are averaged; each signal’s path in the AST is concatenated into a padded vector to capture hierarchical and positional structure.

\subsubsection{\textbf{Clustering}}

Assertions are filtered by structural distance, clustered by semantic features, and then cluster labels are one-hot encoded and fused with PCA-reduced structural vectors for final semantically and structurally consistent grouping.

\subsubsection{\textbf{Specification Split and Functional Points Extraction}}

The specification is split into Sub-SPECs by functional module using an LLM, and fine-grained validation functional points—atomic statements derived directly from the spec—extracted by an LLM for precise coverage feedback.

\subsubsection{\textbf{Assertion-to-Specification Mapping}}

Each assertion group is matched to the most relevant Sub-SPEC, and individual assertions are aligned to specific functional points via signals and descriptions to identify coverage gaps.

\subsubsection{\textbf{Coverage-Driven Feedback Loop}}

Uncovered Sub-SPECs and functional points are fed back to guide iterative assertion generation, focusing on under-verified regions until all Sub-SPECs reach the coverage threshold $\theta=0.85$.

\section{Experiments}

\subsection{Experimental Setup}

Benchmark data is from an open-source dataset \cite{18}, with correctness checked via Cadence JasperGold (v21.12.002). CoverAssert was compared with AssertLLM \cite{10} and Spec2Assertion \cite{14}, all using GPT-4o under identical conditions. We recorded the total generated $N$, syntax-correct $S$, and FPV-passed SVAs $P$, along with branch $BFC$, statement $SFC$, and toggle $TFC$ coverage within each assertion’s Cone of Influence \cite{19}. Four designs of varying scales were selected for evaluation, as summarized in Table \ref{Summary of Design}.

\begin{table}[h]
\centering
\renewcommand{\arraystretch}{1.1} 
\caption{Summary of designs.}
\label{Summary of Design}
\setlength{\tabcolsep}{3.5pt}
\begin{tabular}{c|c|c|c}
\hline
\hline
\cellcolor{gray!20}\textbf{Design Name} & \cellcolor{gray!20}\textbf{Func. Description} & \cellcolor{gray!20}\textbf{LoC} &
\cellcolor{gray!20}\textbf{Num. of Cells} \\ \hline
\makecell{\texttt{I\textsuperscript{2}C}} & \makecell{\hspace*{-4pt} Serial communication protocol.} & \makecell{5369} & \makecell{756}\\
\makecell{\texttt{SHA3}} & \makecell{\hspace*{-17pt} Hash function computation.} & \makecell{141185} & \makecell{22228}\\ 
\makecell{\texttt{ECG}} & \makecell{\hspace*{-12pt} Biological signal acquisition.} & \makecell{398686} & \makecell{59084}\\ 
\makecell{\texttt{Pairing}} & \makecell{\hspace*{-13pt} Cryptographic key exchange.} & \makecell{1561498} & \makecell{228287}\\ 
\hline
\hline
\end{tabular}
\end{table}

\subsection{Experimental Results}

As shown in Table \ref{tab:assert+cover}, we evaluated CoverAssert with AssertLLM and Spec2Assertion on four circuits. $CoverAssert\textbf{-}n$ denotes results after n feedback iterations. The first iteration significantly boosts coverage, and the second further improves most metrics, demonstrating the accuracy of our feedback-guided approach.

\definecolor{deepgreen}{RGB}{110, 251, 152}
\definecolor{lightgreen}{RGB}{188, 251, 152}

\begin{table}[h]
\centering
\caption{Performance comparison of integrating CoverAssert with AssertLLM and Spec2Assertion.}
\label{tab:assert+cover}
\setlength{\tabcolsep}{0.4em} 
\renewcommand{\arraystretch}{0.4} 
\begin{tabular}{@{}>{\setlength{\tabcolsep}{-2em}}l *{5}{c} @{}}
\toprule
\hline
\multicolumn{1}{c}{\cellcolor{gray!20}\textbf{Method}} & \cellcolor{gray!20}\textbf{Metrics} & \cellcolor{gray!20}\textbf{I\textsuperscript{2}C} & \cellcolor{gray!20}\textbf{SHA3} & \cellcolor{gray!20}\textbf{ECG} & \cellcolor{gray!20}\textbf{Pairing} \\
\midrule
\multirow{5}{*}{$\mathrm{AssertLLM}$} & $N$/$S$/$P$ & 127/112/58 & 31/31/26 & 44/44/19 & 32/32/12 \\
 & $BFC(\%)$ & 80.23 & 92 & 82.22 & 76.12 \\
 & $SFC(\%)$ & 82.26 & 90.24 & 80.74 & 83.63 \\ 
 & $TFC(\%)$ & 62 & 78.41 & 57.89 & 67.12 \\
\addlinespace
\multirow{5}{*}{\begin{tabular}{@{}l@{}}$\mathrm{AssertLLM}$ \\ \ \ \ \ \ \ \ $\mathrm{+}$ \\ $\mathrm{CoverAssert\text{-}1}$\end{tabular}} & $N$/$S$/$P$ & 149/133/76 & 69/67/47 & 62/60/31 & 51/49/26 \\
 & $BFC(\%)$ & \cellcolor{lightgreen}85.33 & \cellcolor{lightgreen}98.82 & \cellcolor{lightgreen}98.89 & \cellcolor{lightgreen}90.96 \\
 & $SFC(\%)$ & \cellcolor{lightgreen}87.27 & \cellcolor{lightgreen}94.63 & \cellcolor{lightgreen}97.78 & \cellcolor{lightgreen}93.18 \\ 
 & $TFC(\%)$ & \cellcolor{lightgreen}66.98 & \cellcolor{lightgreen}86.57 & \cellcolor{lightgreen}81.86 & \cellcolor{lightgreen}70.33 \\
\addlinespace
\multirow{5}{*}{\begin{tabular}{@{}l@{}}$\mathrm{AssertLLM}$ \\ \ \ \ \ \ \ \ $\mathrm{+}$ \\ $\mathrm{CoverAssert\text{-}2}$\end{tabular}} & $N$/$S$/$P$ & 173/151/85 & 82/80/56 & 78/76/38 & 64/62/32 \\
 & $BFC(\%)$ & \cellcolor{deepgreen}86.79 & \cellcolor{deepgreen}100 & \cellcolor{lightgreen}98.89 & \cellcolor{deepgreen}93.85 \\
 & $SFC(\%)$ & \cellcolor{deepgreen}88.87 & \cellcolor{deepgreen}95.06 & \cellcolor{lightgreen}97.78 & \cellcolor{deepgreen}94.67 \\ 
 & $TFC(\%)$ & \cellcolor{deepgreen}80.52 & \cellcolor{deepgreen}90.71 & \cellcolor{deepgreen}83.27 & \cellcolor{deepgreen}81.46 \\
\addlinespace
\hline
\hline
\\
\multirow{5}{*}{$\mathrm{Spec2Assertion}$} & $N$/$S$/$P$ & 90/89/49 & 42/42/28 & 35/29/19 & 45/41/15 \\
 & $BFC(\%)$ & 87.87 & 90.89 & 79.51 & 83.58 \\
 & $SFC(\%)$ & 89.44 & 87.78 & 81.05 & 66.67 \\ 
 & $TFC(\%)$ & 64.13 & 65.43 & 73.92 & 46.53 \\
\addlinespace
\multirow{5}{*}{\begin{tabular}{@{}l@{}}$\mathrm{Spec2Assertion}$ \\ \ \ \ \ \ \ \ $\mathrm{+}$ \\ $\mathrm{CoverAssert\text{-}1}$\end{tabular}} & $N$/$S$/$P$ & 105/104/61 & 65/65/46 & 51/42/29 & 76/68/34 \\
 & $BFC(\%)$ & \cellcolor{lightgreen}89.34 & \cellcolor{lightgreen}94.36 & \cellcolor{lightgreen}97.33 & \cellcolor{lightgreen}86.91 \\
 & $SFC(\%)$ & \cellcolor{lightgreen}90.32 & \cellcolor{lightgreen}90.84 & \cellcolor{lightgreen}96.48 & \cellcolor{lightgreen}81.06 \\ 
 & $TFC(\%)$ & \cellcolor{lightgreen}75.06 & \cellcolor{lightgreen}74.03 & \cellcolor{lightgreen}75.26 & \cellcolor{lightgreen}58.35 \\
 \addlinespace
\multirow{5}{*}{\begin{tabular}{@{}l@{}}$\mathrm{Spec2Assertion}$ \\ \ \ \ \ \ \ \ $\mathrm{+}$ \\ $\mathrm{CoverAssert\text{-}2}$\end{tabular}} & $N$/$S$/$P$ & 117/116/67 & 77/77/55 & 68/58/37 & 101/90/51 \\
 & $BFC(\%)$ & \cellcolor{lightgreen} 89.34 & \cellcolor{deepgreen}95.07 & \cellcolor{deepgreen} 98.02 & \cellcolor{lightgreen} 86.91 \\
 & $SFC(\%)$ & \cellcolor{lightgreen} 90.32 & \cellcolor{deepgreen}92.84 & \cellcolor{lightgreen}96.48 & \cellcolor{deepgreen}82.94 \\ 
 & $TFC(\%)$ & \cellcolor{deepgreen}82.42 & \cellcolor{deepgreen}77.75 & \cellcolor{deepgreen}82.85 & \cellcolor{deepgreen}61.96 \\
\hline
\bottomrule
\end{tabular}
\end{table}

\section{Conclusion}

This paper presents CoverAssert, a framework that improves SVA generation via functional coverage feedback. Combining semantic and structural features, it groups and matches assertions, guiding generators to prioritize uncovered points and enhance coverage, while integrating with existing methods.

\section*{Acknowledgment}

This paper is supported in part by the Chinese Academy of Sciences under grant No. XDB0660103, and in part by  the National Natural Science Foundation of China (NSFC) under grant No.( 62090024).

\bibliographystyle{plain} 
\bibliography{reference}

\end{document}